\documentclass[journal,letterpaper]{IEEEtran}
\pdfoutput=1

\usepackage[utf8]{inputenc}
\usepackage[english]{babel}
\ifCLASSINFOpdf
  \usepackage[pdftex]{graphicx}
  \graphicspath{{./figures/}{./bio/}}
  \DeclareGraphicsExtensions{.pdf,.jpeg,.jpg,.png}
\fi
\usepackage[caption=false,font=footnotesize]{subfig}
\usepackage[export]{adjustbox}
\usepackage{mathtools,amssymb,units}
\usepackage[linesnumbered,lined,ruled]{algorithm2e}
\usepackage{booktabs,url}

\begin{document}

\title{Parked Cars are Excellent Roadside Units}
\author{Andre~B.~Reis, Susana~Sargento, and~Ozan~K.~Tonguz
\thanks{
A.~B.~Reis is with the Department of Electrical and Computer Engineering, Carnegie Mellon University, Pittsburgh, PA 15213-3890 USA, and also with the Universidade de Aveiro, Instituto de Telecomunicacoes, Aveiro 3810-193, Portugal (e-mail: abreis@cmu.edu).
}
\thanks{S.~Sargento is with the Universidade de Aveiro, Instituto de Telecomunicacoes, Aveiro 3810-193, Portugal (e-mail: susana@ua.pt).
}
\thanks{O.~K.~Tonguz is with the Department of Electrical and Computer Engineering, Carnegie Mellon University, Pittsburgh, PA 15213-3890 USA (e-mail: tonguz@ece.cmu.edu).
}
}

\maketitle

\begin{abstract}
A comprehensive implementation of the envisioned traffic safety and efficiency applications of the IEEE 802.11p and WAVE standards assume the premise of the use of DSRC technology both as on-board units (OBUs) and as Roadside Units (RSUs). The high cost associated with RSUs, however, has so far prevented massive deployment of RSUs. Finding alternative solutions to this longstanding problem is therefore very important. In this paper, we propose a self-organizing network approach to using parked cars in urban areas as RSUs. This self-organizing network approach enables parked cars to create coverage maps based on received signal strength and make important decisions, such as if and when a parked car should serve as an RSU. Our results show the feasibility and cost-effectiveness of the proposed approach, which is able to provide excellent coverage using only a small fraction of the cars parked in a city.
\end{abstract}

\section{Introduction} 
\label{sec:introduction}

\IEEEPARstart{V}{ehicular} networks require a minimum number of cars to be well connected and functional, which can fail to happen due to either low numbers of vehicles on the road or insufficient radio-equipped vehicles. Studies show that in areas of low vehicle density, important safety broadcasts can take more than 100~seconds to reach all nearby cars~\cite{uvcast}. On the other hand, dense networks with too many vehicles can be overwhelmed with traffic and signaling~\cite{eichler2007}, requiring careful coordination between the nodes to ensure proper operation.

One way to overcome both these problems is to supplement vehicle-to-vehicle (V2V) communications with vehicle-to-infrastructure (V2I) support systems, that is, by deploying infrastructure nodes known as Road-Side Units (RSUs) along the road, in addition to the Dedicated Short Range Communications~(DSRC)~/~IEEE 802.11p units within the vehicles~\cite{ieee80211p}. These units can supplement a sparse network in a low-density scenario, and help coordinate and move data in dense scenarios. 

The U.S. Department of Transportation (DoT) anticipated a nationwide deployment of supporting infrastructure of RSUs for vehicular networks to have happened by 2008~\cite{dotfreitas} -- however, this forecast did not come to fruition due to difficulties in justifying the benefits of RSUs, lack of cooperation between the public and private sectors, but most importantly, a lack of funding for infrastructure whose widespread deployment was estimated to cost billions of dollars. A 2012 industry survey by Michigan's DoT and the Center for Automotive Research reiterated that ``\emph{one of the biggest challenges respondents see to the broad adoption of connected vehicle technology is funding for roadside infrastructure}.''~\cite{michiganstudy} and, in 2014, a nationwide study sponsored by the U.S.~DoT reported average costs of \$17,680 per deployed Roadside Unit~\cite{dotcosts}, for both hardware and installation. These prohibitive costs explain why one is unlikely to see substantial deployments of RSUs, despite their importance to vehicular networks.

One way to avoid the expense of an infrastructure deployment is to use the vehicles themselves as RSUs~\cite{carsasrsus2013}, and in urban areas the cars that are parked can also be leveraged to serve as RSUs~\cite{vtc2015}. The objective of this paper is to propose a credible, low-cost alternative to a Roadside Unit deployment that can operate both independently and in conjunction with existing RSUs. 

To this end, we introduce \emph{a self-organizing network approach} that allows parked cars to function as RSUs forming a vehicular support network. We tackle the problem of selecting which parked cars should become part of the network, how to measure each car's utility to be able to make informed decisions, what algorithmic steps should vehicles follow when they park, and how to deal with possible disruptions in connectivity when parked cars leave.

Simulation data from a unique platform with real-life data validate our proposed approach and its methods. The results show that an on-line, greedy decision process based on self-generated coverage maps and limited communication between vehicles can achieve an average coverage that is 93\% to~97\% as good as an optimum solution. The proposed algorithms can also optimize the number of parked cars that are designated as RSUs, selecting on average only 12\% more cars than an optimum process. A second set of data shows, additionally, that parked cars bring tangible benefits in initial deployment stages, where insufficient numbers of DSRC-enabled vehicles cause the network to become sparse. In these scenarios, using small numbers of parked cars to act as RSUs can reduce the time for emergency messages to be broadcast by 40-50\%. Through these key results, we verify that parked cars can indeed serve both as RSUs and as an extension to vehicular infrastructure deployments, and can self-organize to do so with no significant overhead to the existing networks of moving cars.

\begin{figure*}[t]
	\centerline{
		\subfloat[]{\includegraphics[trim=6cm 4cm 6cm 4cm,clip,frame,width=0.31\linewidth]{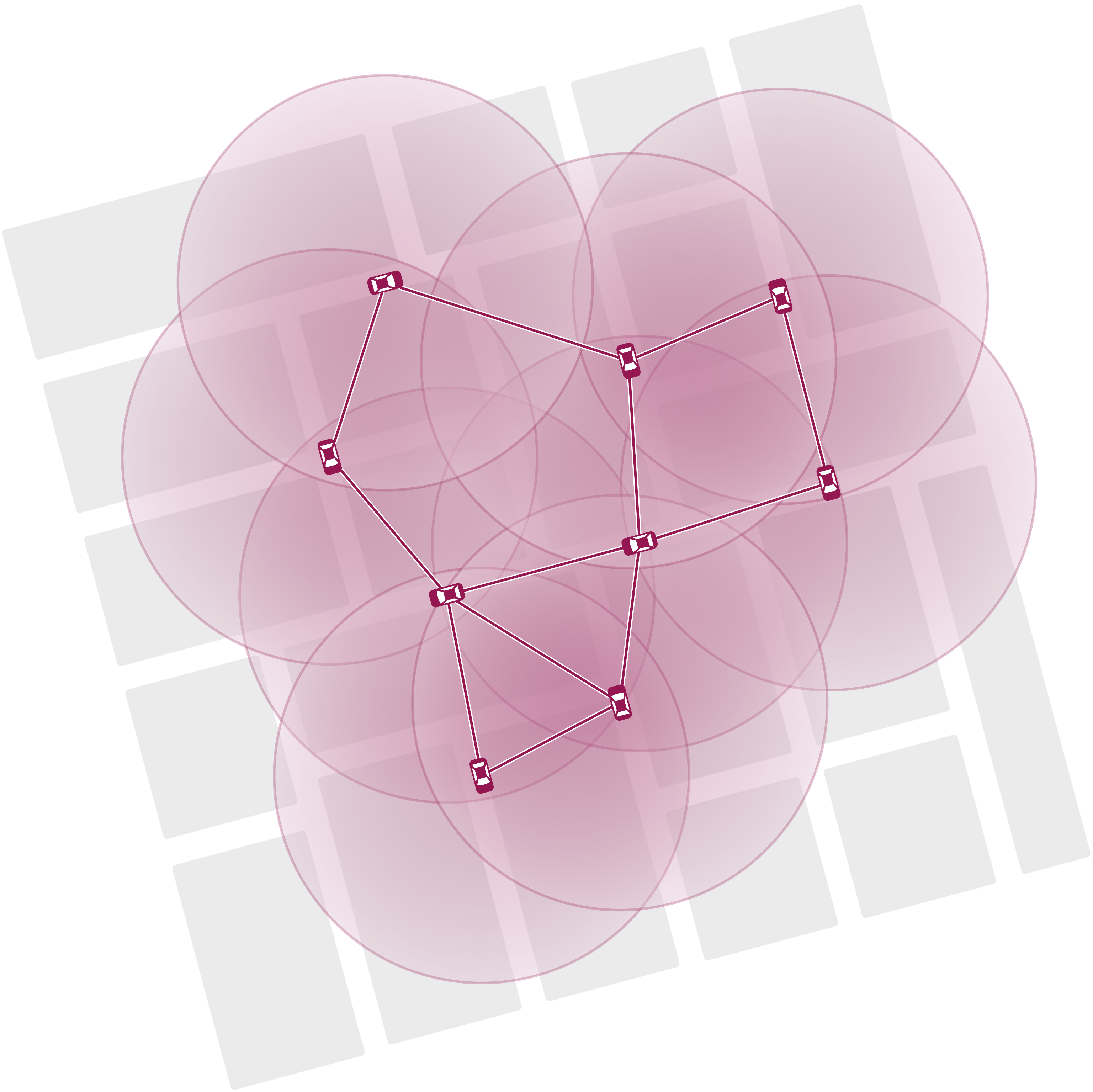}
		\label{fig:modesmesh}} 
			\hfil
		\subfloat[]{\includegraphics[trim=6cm 4cm 6cm 4cm,clip,frame,width=0.31\linewidth]{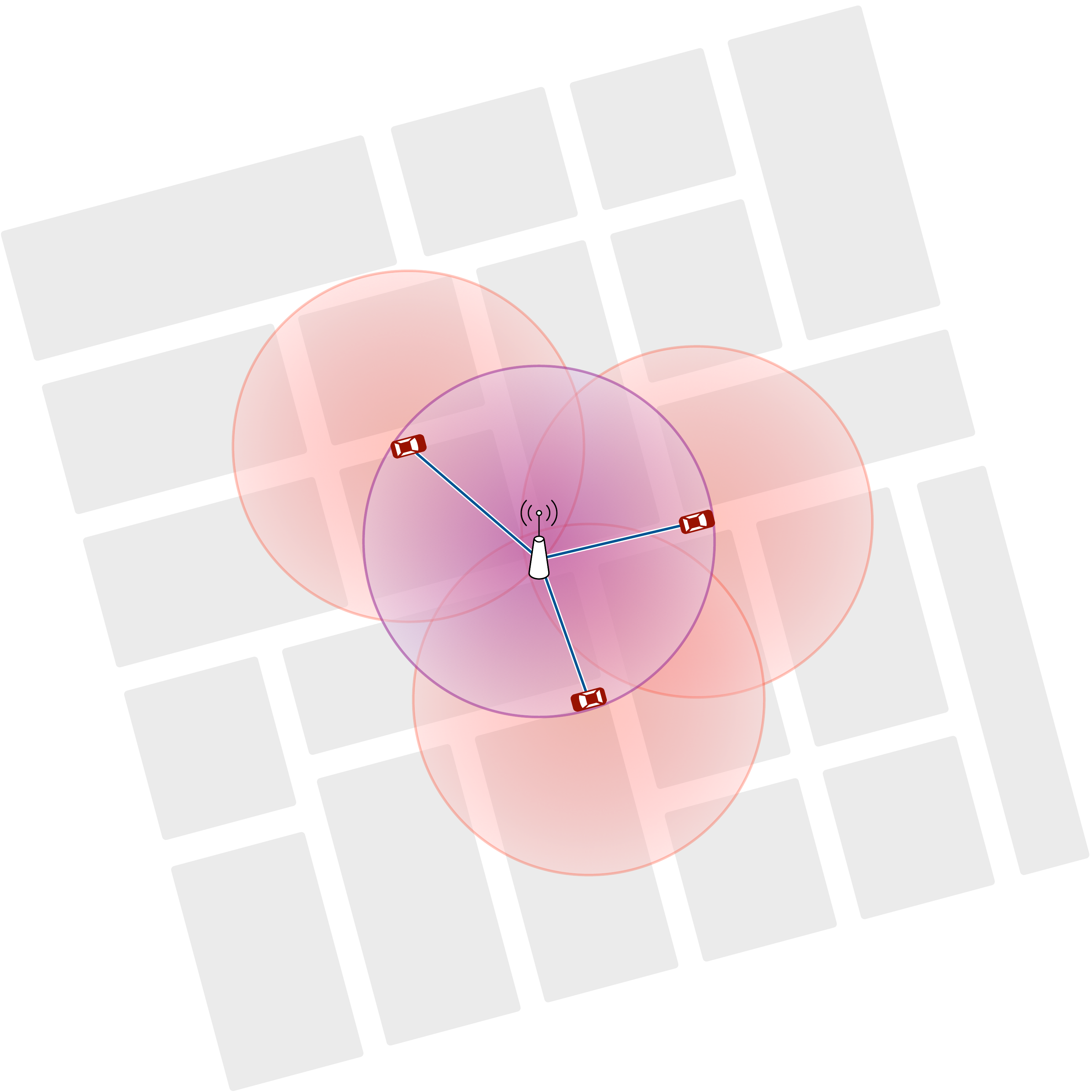} 
		\label{fig:modesrelay}}
			\hfil
		\subfloat[]{\includegraphics[trim=6cm 4cm 6cm 4cm,clip,frame,width=0.31\linewidth]{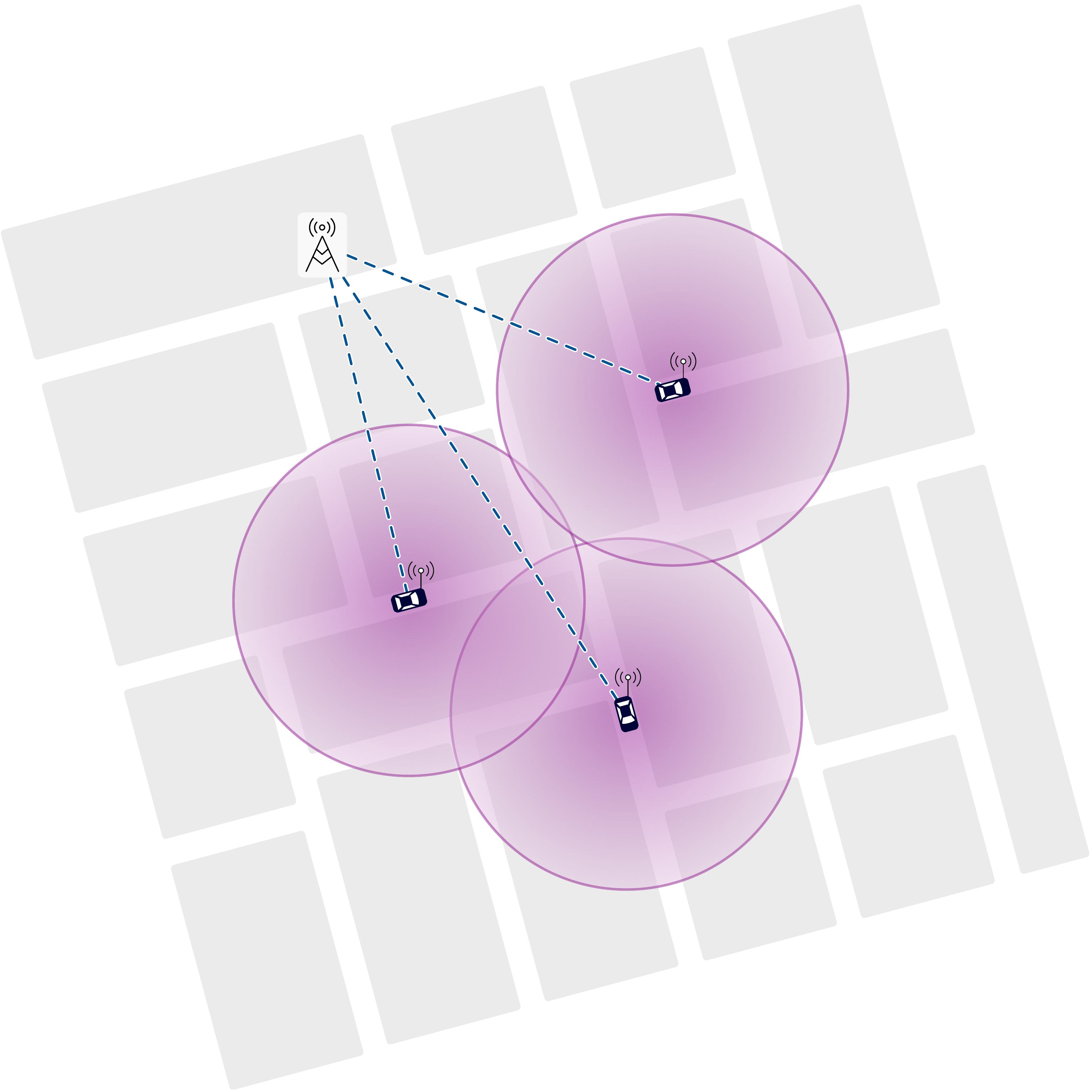} 
		\label{fig:modesstandalone}}
	} 
	\caption{Modes of operation for parked cars acting as RSUs. (a) Parked cars form a mesh network with point-to-point links to other parked cars in range, using 802.11p. (b) Parked cars extend the range of a fixed 802.11p RSU, acting as relays to it. (c) Parked cars with access to an uplink establish themselves as standalone RSUs (depicted: using a cellular network as an uplink).} 
	\label{fig:modesofoperation}
\end{figure*}

This work's main contributions can be summarized as follows.
\begin{itemize}
	\item A low-cost solution to the problem of deploying RSUs for providing support to urban vehicular networks is proposed, by leveraging the parked cars in cities.
 	\item A self-organizing network approach to selecting RSUs from a large pool of parked cars is formulated. We introduce a novel way for vehicles to assess their value to the network by listening to beacons transmitted by other cars on the road, and piecing together a map of their own coverage. 
	\item A simple on-line algorithm is designed that maximises the coverage of the support network of parked cars, while minimizing the number of cars that are required to be enabled. Using the coverage maps, this algorithm requires only brief 1-hop exchanges between RSUs. 
	\item A detailed study is provided for the benefits of the proposed approach at the initial stages of implementation, where the small market penetration rate of DSRC-equipped vehicles will imply that only a few parked cars in a given area will be able to become RSUs.
\end{itemize}

The remainder of this paper is organized as follows. Section~\ref{sec:leveraging_parked_cars_as_roadside_units} gives an overview of the roles parked cars can take in an urban network. Section~\ref{sec:selforganizing} describes the self-organizing approach that is at the core of this work. Section~\ref{sec:performance_analysis} begins with a description of our simulation platform, which is first followed by a study of the advantages this approach can bring to urban areas in initial stages of deployment, and afterwards with an analysis of the self-organization process and how it fares against optimal decisions. Section~\ref{sec:considerations_on_vehicle_battery_life} provides a discussion on the impact of our approach on a vehicle's battery life. Related work is presented in Section~\ref{sec:relatedwork}, and finally, concluding remarks are given in Section~\ref{sec:conclusion}.


\section{Leveraging Parked Cars as Roadside Units} 
\label{sec:leveraging_parked_cars_as_roadside_units}
While the first suggestions for the use of cars as Roadside Units aimed at recruiting the moving vehicles for this purpose~\cite{carsasrsus2013}, in urban areas parked cars present themselves as equally compelling candidates. For one, parked cars share many of the benefits of fixed RSUs, of which one of the most significant is that \emph{they do not move}.

As cars move, buildings and other obstructions enter and leave the line of sight between them, causing detrimental fading and shadowing effects on the communication channel. Eventually, the nodes move too far apart from each other, and the channel disappears. A parked car in an urban area has a fixed, known position for extended periods of time and consequently, a more stable communication channel with nearby cars and RSUs is possible. Their fixed location is also beneficial to applications that rely on \emph{geocasting} (the broadcasting of messages to specific geographic areas), making it simpler to route such messages to their intended location.

We introduce three methods of operation for parked cars to become a part of the vehicular network -- they are depicted in Figure~\ref{fig:modesofoperation}. When no fixed RSUs exist in the urban area, parked cars can form a \textbf{mesh network} with each other (see Figure~\ref{fig:modesmesh}), each parked car connecting to its neighboring parked cars. This mesh network can support the moving vehicles by offering to relay and broadcast messages, with the benefit of having a stable node structure and stable links between nodes (moving cars do not enjoy this benefit).


When a limited deployment of fixed RSUs is already present, parked cars in the vicinity of an RSU can act as \textbf{relays} to it (see Figure~\ref{fig:modesrelay}), extending the reach of the fixed unit. In this mode, the value of a deployed RSU can be increased by enlarging its area of coverage. Parked cars acting as relays advertise their ability to forward messages to an RSU, so moving vehicles use them as such. Around these relays, other parked cars can continue to form the previously-described mesh network, further amplifying the coverage of the original RSU via relays of relays.

Lastly, a parked car with the access to a backbone uplink can leverage that link to establish itself as a \textbf{standalone RSU} (see Figure~\ref{fig:modesstandalone}). In this mode, a parked car can communicate with other existing RSUs and uplink-enabled cars via the Internet, allowing it to send messages to distant locations in adequate time. Increasing numbers of vehicles now come equipped with Internet connectivity (using 3G or LTE/LTE-A cellular technologies), and there is the possibility for the car's electronics to be allowed the use of this link for selected purposes (e.g., safety messages).

In urban areas that have seen limited deployments of fixed infrastructure, parked cars can be used to fill in the gaps, providing coverage in areas not serviced by existing RSUs, directing messages to the RSUs as necessary. Parked cars can, therefore, serve both as the main solution to providing support to a mobile vehicular network and as a complimentary solution to the existing infrastructure. 

The remainder of this work will focus on the crucial issue of deciding which parked cars should become a part of this vehicular support network.


\section{Proposed Solution: A Self-Organizing Network Approach to Creating RSUs} 
\label{sec:selforganizing}
This section describes a self-organizing approach for constructing a vehicular support network from parked cars. Together with the simulation data on sparse networks and the performance analysis on dense networks, these constitute the main research presented in this paper.

At the core of this self-organizing approach lies a single decision: when a vehicle parks, should it become an RSU, or should it enter a sleep mode? The decision depends on what the support network aims to accomplish. When a parked car takes a Roadside Unit role, this has non-negligible costs both to the battery of that vehicle, as it must then power the DSRC electronics, and to the vehicular network, as active nodes will broadcast their presence and answer requests from other cars, causing overhead. So, for dense areas, the primary goal should be to maximize the reach of this support network, while minimizing the number of active RSUs. 

With this objective in mind, we introduce an \emph{on-line}, \emph{greedy} algorithm that allows each car to decide whether to become an RSU or not. The process is necessarily on-line since the network is in constant evolution, as cars park and leave; our approach is also greedy, in the sense that we attempt to make a locally optimum decision whenever a vehicle is parked, aiming to approach a global optimum solution.

With full knowledge of each parked car at every location, one can evaluate all possible combinations of turning cars on and off, and reach a global optimum solution. Evidently this is not feasible, as not only it would quickly amount to millions of computations, it would also need to repeat itself whenever a car parked or left. With this work, we aim for an approach that requires \emph{minimal communication between nodes}.


\subsection{Self-Observed Coverage Maps} 
\label{sub:self_observed_coverage_maps}
To be able to optimize the number of parked cars that become active and take the role of RSUs, we require a new metric that can represent each vehicle's value to the network. The primary goal of this RSU network is to be able to reach as many locations in the city as possible -- therefore, we are interested in knowing the \emph{signal coverage} of each parked car, i.e., which areas it can reach, and how well it can do so. 

Cars, however, can park in numerous distinct locations. Using propagation models, one could estimate the coverage map of each particular car, but doing so would also require the local roadmap, as well as knowing the shapes and structures of nearby buildings (to determine obstructions). Such a process would still fail to account for undocumented obstacles such as trees, trucks, billboards, etc. 

With this work, we introduce a system whereby parked cars listen to beacons being broadcast by other nearby cars on the road and use those beacons to build maps of their coverage. By doing so, parked cars learn about which areas in their vicinity they can send and receive messages from. These beacons, which include position, speed, and bearing, are standardized and known as Cooperative Awareness Messages (CAMs), and are broadcast at rates no lower than 1~Hz~\cite{etsiCAM}. When the signal strength of incoming beacons is made available from the lower protocol layers, parked cars can take advantage of this valuable information to also track how strong coverage is at each location.

\begin{figure*}[t]
 \includegraphics[width=\linewidth]{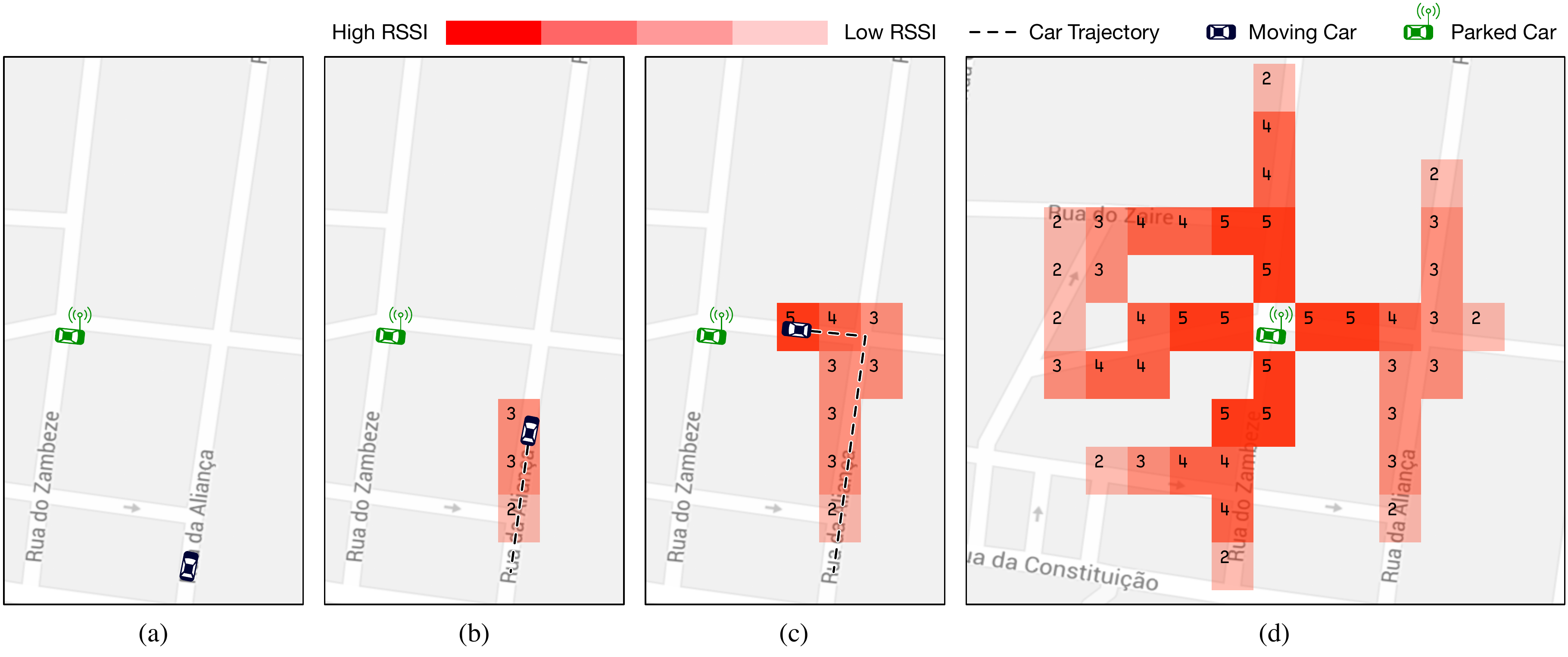}
 \caption{In this example, Figures~\ref{fig:learncoverage}a, \ref{fig:learncoverage}b, and~\ref{fig:learncoverage}c show how a parked car (the green car in the figure) observes parts of its coverage map by listening to beacons from a nearby moving vehicle. As a vehicle moves nearer, our parked car overhears its beacons and corresponding signal strength, tagging the cells in its local map. Figure~\ref{fig:learncoverage}d shows the complete, observed local coverage map. }
 \label{fig:learncoverage}
\end{figure*}

In Figure~\ref{fig:learncoverage}, we illustrate the process of learning a coverage map, while also demonstrating coverage tracking in the shape of cells. In this work, we divide the urban area into a logical 2D cell map that is common to all cars. This approach stems from the need to have coverage maps that are easily representable, storable and shareable, and that can track coverage in irregular street maps with adequate precision. To ensure that all vehicles share the same grid division, we opt to align cell boundaries to GPS coordinates, a universal reference for all GPS-enabled cars.

Under this cellular division, a self-observed coverage map can be represented as a matrix containing signal strength values, with $5$ meaning \emph{excellent signal} and $0$ meaning \emph{no coverage}. In this work, a typical coverage map is $11\times 11$ cells; $3$ bits per cell can represent the $6$ degrees of coverage, which makes a map $\approx 46$~bytes in size. To center the map, the latitude and longitude of the vehicle ($2$~bytes each) is sent along with the data matrix. In this matrix form, coverage maps can be shared between vehicles through the WAVE Short Message Protocol (WSMP), a part of the WAVE standard~\cite{wave16093}.



\subsection{Procedures for Newly Parked Cars} 
\label{sub:procedures_for_newly_parked_cars}
\begin{figure}[t]
 \centering
 \includegraphics[width=\linewidth]{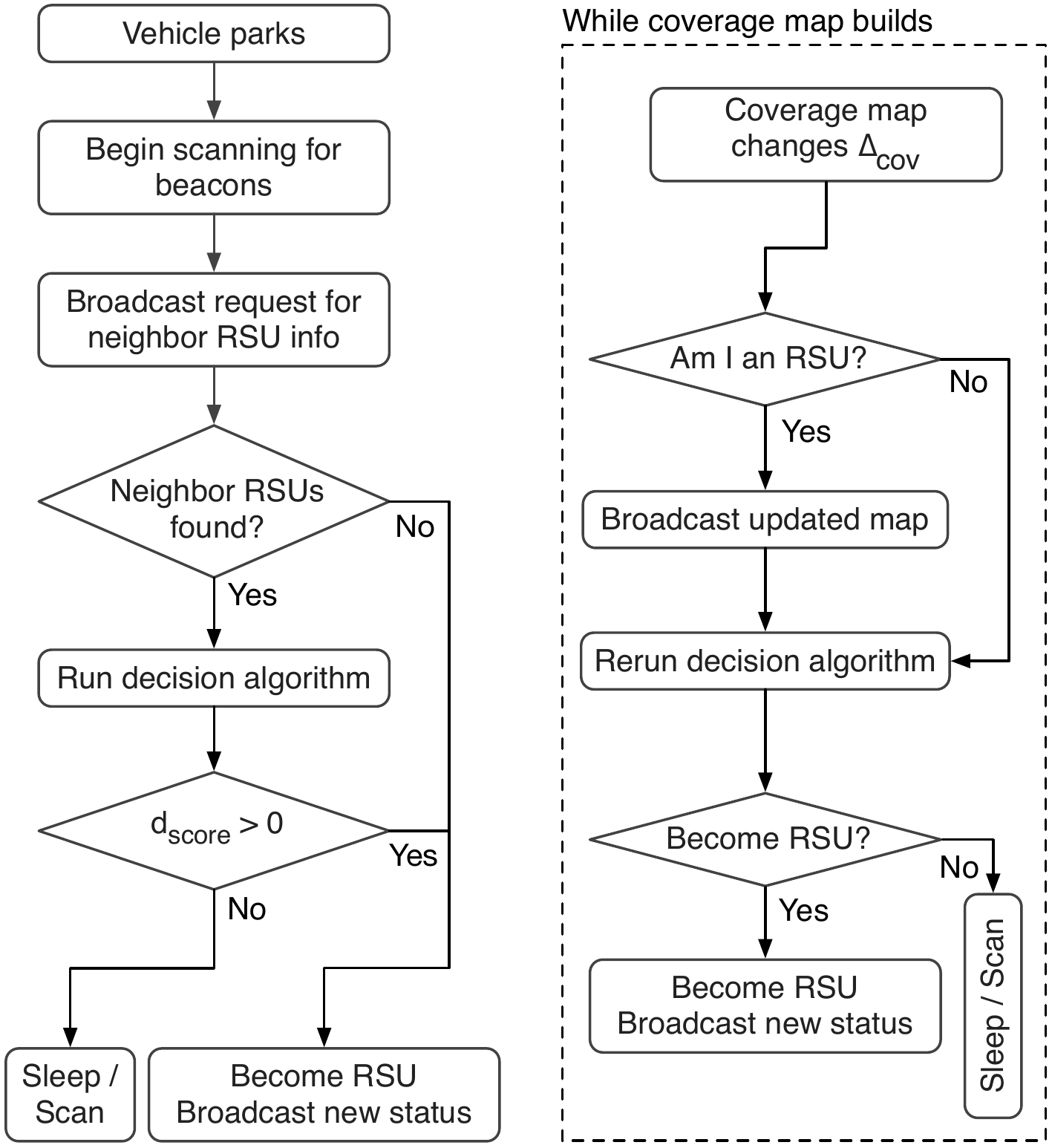}
 \caption{(a) Flowchart of actions taken by a newly parked car in order to decide whether to become an RSU. (b) A parked car may see a substantial update to its coverage map, triggering a new decision process.}
 \label{fig:flowchart}
\end{figure}

Each newly parked car must follow a logical set of actions to allow it to decide whether it should become an RSU. A newly parked car begins by listening for beacons being broadcast by other vehicles in its vicinity, as seen in Figure~\ref{fig:flowchart}a. During this process, the vehicle builds its coverage map, as described in the last section. We provide reference values for the duration of this step later on, in Section~\ref{sub:performance_of_self_organization_with_minimal_information}. 

Once the step of building a map of coverage is complete, the vehicle requests the coverage maps of neighboring active RSUs, using WAVE Short Messages (WSMs). The decision process only requires maps from fixed RSUs and active parked cars. An alternative to this step is to instruct RSUs to broadcast their coverage maps periodically, so newly parked cars can simply receive them while listening for beacons.

The outcome of this algorithm ultimately decides whether the parked car should become an RSU or switch to a power-saving (sleep) mode. The decision made by each car depends on its observed coverage and the maps it receives from other RSUs in its 1-hop neighborhood. We opt to keep the decision down to each vehicle, while relying only on information from the car's 1-hop neighbor nodes, to minimize the associated network overhead. We describe this decision process in more detail in the next section.

The process that newly parked cars must follow is straightforward. However, it is possible for a car's coverage map to change after the decision step, which can occur if no vehicles passed through a nearby road during the listening step. Coverage can also worsen if an obstruction appears near a parked car, e.g. if a large truck decides to park next to it. To account for this possibility, when a car detects that its coverage map has changed significantly, it reevaluates its decision, by following the steps in Figure~\ref{fig:flowchart}b. A delta value, $\Delta_{cov}$, tracks how many cells have changed since the last decision, and crossing a prespecified threshold triggers a new decision step. 



\subsection{Making a Decision} 
\label{sub:making_a_decision}
At the core of the decision process lies the \emph{decision score} ($d_{score}$), a measure of each car's added value to the network. As mentioned in the previous section, each car must make a decision when it parks, once it has inferred its coverage map and received the maps from nearby RSUs (which can be other active parked cars, as well). With this data in hand, the now-parked vehicle estimates what will happen to the network should it decide to become an RSU at the location where it parked. In essence, it measures its net benefit against its perceived detriment. 

The coverage matrixes from neighboring RSUs are combined to obtain two maps of the local area: a first map with the best signal coverage avaliable at every cell, and a second map with the number of RSUs that serve each cell. When the decision process is triggered, the new parked car overlays its coverage map with these local area maps. For each cell in this map, the parked car will see one of the following occur:
\begin{itemize}
 \item \textbf{Establishing new coverage.} If the new parked car is covering a cell that no other RSU in its vicinity can reach, then becoming an RSU will provide new coverage to the network. This action is a major benefit.
 \item \textbf{Improving existing coverage.} If the newly parked car can cover an existing cell at a signal level better than any other nearby RSU, then becoming an RSU will improve the strength of the existing coverage provided by the RSUs. This action also benefits the network.
 \item \textbf{Adding redundant coverage.} A parked car that adds coverage to an already-covered cell, but does so at a signal strength level equal to or lower than what is already provided by other RSUs, is adding unnecessary weight to the network. We refer to this as \emph{saturating} the network.
\end{itemize}

Each of these three effects is then quantified, and a weighted sum of each results in the decision score. In this work, $d_{score}$ is given by 
\begin{equation}
 \mathrm{d_{score}} = \kappa \cdot \mathrm{d_{new}} + \lambda \cdot \mathrm{d_{boost}} - \mu \cdot \mathrm{d_{sat}} \quad ,
 \label{eq:dscore}
\end{equation}
where $\mathrm{d_{new}}$, $\mathrm{d_{boost}}$ and $\mathrm{d_{sat}}$ are metrics for \emph{new coverage}, \emph{improved coverage}, and \emph{excess coverage}, respectively. The coefficients $\kappa$, $\lambda$ and $\mu$ weight each component of the decision score, adjusting the balance between improvement and degradation to the network. Later on, in Section~\ref{sub:performance_of_self_organization_with_minimal_information}, we will provide reference values for these coefficients.

\newcommand{\SCM}{\mathrm{\textsc{scm}}}
\newcommand{\LMC}{\mathrm{\textsc{lmc}}}
\newcommand{\LMS}{\mathrm{\textsc{lms}}}
\newcommand{\scm}{\mathrm{scm}}
\newcommand{\lmc}{\mathrm{lmc}}
\newcommand{\lms}{\mathrm{lms}}

We now formalize each of these metrics and how they are calculated. All notation is summarized in Table~\ref{tab:notation}. 
\begin{table}[b]
\centering\small
\caption{Notation Reference}
\label{tab:notation}
\begin{tabular}{@{}p{4cm}l@{}}
\toprule
Definition & Notation \\ 
\midrule
Matrix indices (row, col.) & $i,j$ \\
Geographic indices (lat., lon.) & $x,y$ \\
Self-observed coverage map & $[\SCM] = (\scm_{ij})$ \\
Center coordinates of an $\SCM$ & $x_{center},y_{center}$ \\
Deciding vehicle's own $\SCM$ & $\SCM_0$ \\
Colection of neighbor $\SCM\mathrm{s}$ & $\mathcal{N} = \{ \SCM_1, \SCM_2, \ldots, \SCM_n \}$ \\
Local map of coverage & $[\LMC] = (\lmc_{ij})$ \\
Local map of saturation & $[\LMS] = (\lms_{ij})$ \\
\bottomrule
\end{tabular}
\end{table}
A newly parked vehicle on its decision step has its own observed coverage map, $[\SCM_0]$, whose elements $\scm_{ij}$ represent geographic cells.
At this point, the vehicle also holds a collection of its neighbors' coverage maps, $\mathcal{N} = \{ \SCM_1, \SCM_2, \ldots, \SCM_n \}$. Coverage matrixes are square matrixes of order $n$ (but not symmetric, i.e., $\scm_{n1}\neq{}\scm_{1n}$) where $n$ is always odd, so that a single cell exists at the center of the matrix that corresponds to the vehicle's location.

In this work, we use the subscripts~$({i,j})$ to refer to matrix rows and columns, and the subscripts~$({x,y})$ to refer to geographic coordinates. Coverage matrixes are always accompanied by the coordinates~$(x_{center},y_{center})$ of the center cell $\scm_{\lceil{}n/2\rceil\lceil{}n/2\rceil}$, which allow any element to be mapped to a latitude/longitude pair -- when $({x,y})$ is used, the mapping is implied, since it is straightforward.

We begin by constructing two cell maps of the local neighborhood: a local map of signal coverage $\LMC = (\lmc_{ij}) \in \{0,1,\ldots,5\}$ and a local map of saturation (RSU redundancy) $\LMS = (\lms_{ij}) \in \mathbb{N}_0$. These maps give a picture of the existing RSU support network: $\LMC$ tracks the best coverage being provided at each cell, and $\LMS$ counts the number of RSUs covering that cell at the same time. Algorithm~\ref{alg:localmaps} shows how these local maps are built.
\begin{algorithm}
\DontPrintSemicolon
\KwData{$\mathcal{N} = \{ \SCM_1, \SCM_2, \ldots, \SCM_n \}$}
\KwResult{Local maps $\LMC$, $\LMS$}
$\triangleright$ $\lmc_{xy}$ and $\lms_{xy}$ are initialized to $0$ \;
\ForEach{$\SCM_n\in\mathcal{N}$}{
 \ForEach{$\scm_{n[xy]}\in\SCM_n$}{
 \lIf{$\scm_{n[xy]} > \lmc_{xy}$}{$\lmc_{xy}\gets{}\scm_{n[xy]}$}
 \lIf{$\scm_{n[xy]} > 0$}{$\lms_{xy}\gets{}\lms_{xy}+1$}
 }
}
\caption{BuildLocalMaps\label{alg:localmaps}}
\end{algorithm}

Local maps span from the lowest to the highest latitudes/longitudes seen in all $\SCM$s, i.e., they are as large as the total geographic area covered by the underlying self-observed coverage maps. 

With the $\LMC$ and the $\LMS$, the metrics in Equation~\eqref{eq:dscore} can now be calculated as described in Algorithm~\ref{alg:scoremetrics} below. $\mathrm{\mathbf{d_{new}}}$ sums the strength of new coverage: e.g., a new cell covered with signal `4' will add `4' to $\mathrm{d_{new}}$. This way, the decision process can distinguish between new coverage that can be stronger or weaker. $\mathrm{\mathbf{d_{boost}}}$ measures the improvements to already-covered cells, summing the delta between the existing and new signal levels. $\mathrm{\mathbf{d_{sat}}}$ tracks redundant coverage (network saturation): for every cell that the new vehicle covers, sum the number of neighbors that already service it. For example, if a given cell that the car can reach is already serviced by 5 RSUs, add `5' to $\mathrm{d_{sat}}$. This additive increase process helps keep a balanced number of RSUs that serve each cell. 

\begin{algorithm}
\DontPrintSemicolon
\KwData{$\SCM_0$, $\LMC$, $\LMS$}
\KwResult{A tuple of score metrics $(\mathrm{d_{new}},\mathrm{d_{boost}},\mathrm{d_{sat}})$}
$\mathrm{d_{new}} \gets \mathrm{d_{boost}} \gets \mathrm{d_{sat}} \gets 0$ \;
\ForEach{$\scm_{xy}\in\SCM_0$}{
 \If{$\scm_{xy}>0$}{
 \uIf{$\lmc_{xy}=0$}{
 $\mathrm{d_{new}} \gets \mathrm{d_{new}} + \scm_{xy}$
 }
 \ElseIf{$\lmc_{xy} < \scm_{xy}$}{
 $\mathrm{d_{boost}} \gets \mathrm{d_{boost}} + (\scm_{xy} - \lmc_{xy})$ 
 }
 $\mathrm{d_{sat}} \gets \mathrm{d_{sat}} + \lms_{xy}$
 }
}
\caption{ScoreMetrics\label{alg:scoremetrics}}
\end{algorithm}

\begin{figure}[!t]
 \centering
 \includegraphics[width=1.00\linewidth]{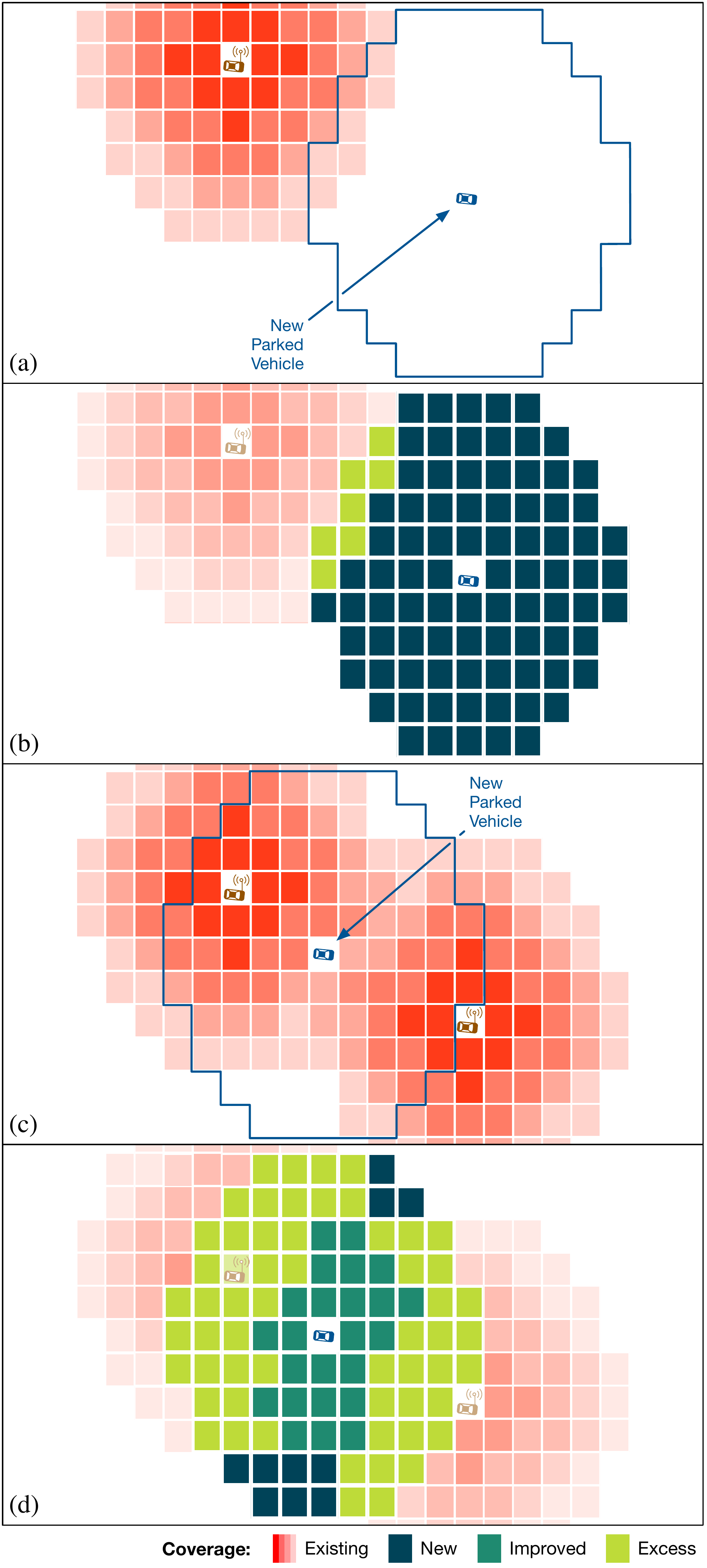}
 \caption{
 A new vehicle parks in an advantageous location (Figure~\ref{fig:dscore}a), in the vicinity of a single existing RSU. The decision algorithms (Figure~\ref{fig:dscore}b) show that it can cover a substantial new area, and so the car establishes itself as a second RSU.
 Later on, another vehicle parks in between the two existing RSUs (Figure~\ref{fig:dscore}c). The algorithms now show (Figure~\ref{fig:dscore}d) that this car would be a net loss to the network, adding more redundant cells than new and improved cells. This vehicle does not become an RSU.
 }
 \label{fig:dscore}
\end{figure}

Figure~\ref{fig:dscore} displays two contrasting examples of the self-organizing decision process. The first, in Figures~\ref{fig:dscore}a and~\ref{fig:dscore}b, shows a new car parking in an advantageous location and evaluating its decision score. The algorithms reveal that, of the 93 cells in the car's coverage map, 86 will add new coverage to the network while 7 will bring no improvement to the existing coverage. This car will, therefore, have a positive~$\mathrm{d_{score}}$ and become an RSU. 
With two RSUs now active, Figures~\ref{fig:dscore}c and~\ref{fig:dscore}d give a second example where another vehicle parks in between the RSUs. This new vehicle sees, from its decision algorithms, that it can add coverage to 10 new cells, and improve existing coverage to 25 cells -- however, it will also be adding to network saturation to 58 cells, and may therefore have a negative~$\mathrm{d_{score}}$, and enter a sleep state.

The decision process builds from coverage matrixes that reflect the real-life status of the RSU support network and returns a Yes/No decision on each newly parked car. With the procedure for gathering coverage maps, the algorithms for newly parked cars, and the decision score equation, the self-organizing approach is fully functional. 



\subsection{Substituting Displaced Cars} 
\label{sub:substituting_displaced_cars_in_a_network_friendly_way}
We now outline an optional mechanism to preserve the structure of the support network whenever an RSU is displaced. When a parked car that is taking a Roadside Unit role becomes mobile again, we refer to this situation as the RSU having been \emph{displaced}. Because decisions only occur when cars park, the service that was being provided by that RSU is not reestablished until a new car parks in the area.

One optimization to the self-organizing approach is to allow inactive parked cars to wake up periodically and listen to the 802.11p Control Channel (CCH), so they can react to changes in the network or respond to requests to come back on-line. We will first show how to achieve this periodic wake-up, and then propose a limited communication system for inactive vehicles to elect a replacement to a displaced RSU.

\subsubsection{Periodic Wake-Up} 
\label{ssub:periodic_wake_up}
A periodic wake-up allows inactive parked cars to be recalled in case of need. For a replacement protocol to be able to act upon all viable candidates, the wake-up events must be synchronized, which can be achieved with ease through 802.11p's mandatory time synchronization. 

Through Time Advertisement messages and GPS timecode data, all On-Board Units (OBUs) of a vehicular network are synchronized to UTC (Universal Time Coordinated). We can then implement modular arithmetics based on the local time $t_{\mathrm{OBU}}$ to ensure all OBUs wake up in parallel. A modulo operation triggers a wake-up when the remainder of $(t_{\mathrm{OBU}}~\mathrm{mod}~N)$ is zero: e.g., with $t_{\mathrm{OBU}}$ in seconds and $N=15$, the OBUs will wake 4 times a minute. On each wake-up, the OBU must listen for a single CCH interval (50~ms), so with a 15-second interval inactive cars to have their OBUs active 0.3\% of the time, with a corresponding energy savings. There is therefore a tradeoff between the reaction time of inactive parked cars and the energy required to allow inactive cars to be contacted. 
 

\subsubsection{Electing a Replacement} 
\label{ssub:electing_a_replacement}
\begin{figure}[t]
 \centering
 \includegraphics[width=\linewidth]{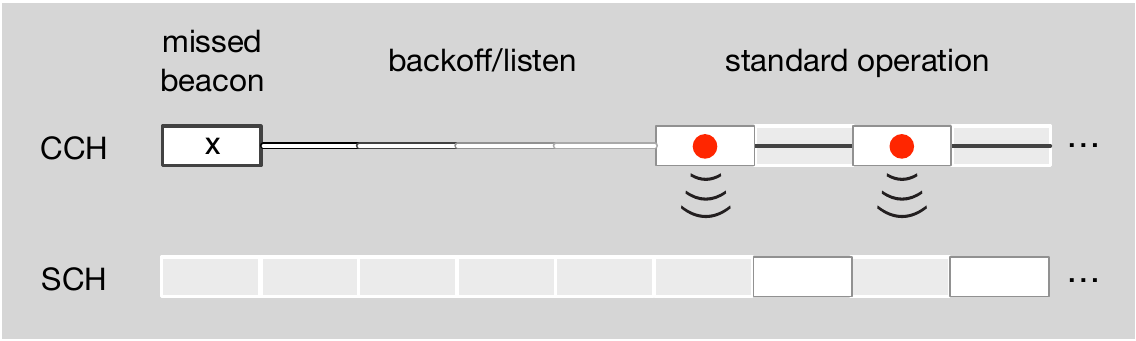}
 \caption{The proposed election process, occuring on every inactive car that detects the loss of an RSU. This process occurs exclusively in IEEE~1609.3's Continuous Channel Access, where the radios remain on the Control Channel (CCH), and do not switch to the Service Channels (SCH). Slots correspond to CCH Intervals, standardized as 50~ms.}
 \label{fig:missedbeacon}
\end{figure}

An inactive parked car detects a displaced RSU when it fails to hear a beacon coming from said RSU during its periodic wake-up interval. Beacons in 802.11p do not benefit from guaranteed delivery, so a car should only react once it fails to hear multiple beacons in a row. Figure~\ref{fig:missedbeacon} illustrates the election process, which works by quickly eliminating as many candidates as possible with no communication between them. This silent elimination ensures that a network in an area with hundreds of inactive cars (e.g. in a parking lot) is not flooded with messages every time an RSU is displaced.

To this end, we propose an initial backoff period inversely related to each car's decision score, as described earlier in Section~\ref{sub:making_a_decision}. Each car detecting a displaced RSU computes its $\mathrm{d_{score}}$ setting aside the coverage map of the displaced RSU. Its backoff time $t_{\mathrm{backoff}}$ is given by
\begin{equation}
 t_{\mathrm{backoff}} = 
 \left\lfloor 
 \left( 1-
 \frac{ \mathrm{d_{score}} }{ \mathrm{d_{score_{max}}} } \right) 
 \times 
 N_{\begin{subarray}{l}\textrm{backoff}\\\textrm{slots}\end{subarray}} 
 \right\rfloor 
 \times 
 t_\mathrm{CCH} \quad ,
 \label{eq:backoff}
\end{equation}
where $\mathrm{d_{score_{max}}}$ is the largest value a decision score can take (which is found empirically), $N_{\mathrm{backoffslots}}$ specifies how many intervals the backoff process is allowed to last for, and $t_{\mathrm{CCH}}$ is the duration of a CCH Interval, 50~ms. In this calculation, the car's decision score is used as a ratio to the number of backoff intervals. Larger (better) decision scores will then lead to shorter wait times. As the backoff timer on the best candidate expires, it begins to broadcast new RSU beacons, advertising itself as the winner of the process and instructing other candidates to return to sleep. 

The number of backoff slots determines the balance between how quickly a replacement RSU is found and how many candidates are excluded by the process. When multiple listeners are assigned the same expiration timer, contention at the Medium Access layer ensures that only one will broadcast a beacon first, but at this point, the choice is random. $N_{\mathrm{backoffslots}}$ should therefore be sufficiently large to eliminate most candidates based on their decision scores (e.g., 40 backoff slots can exclude 97.5\% of all candidates in 2 seconds).




\section{Performance Analysis} 
\label{sec:performance_analysis}
Here, we present a study of the benefits that parked cars can bring to sparse urban areas and an analysis of the performance of the self-organizing approach shown in Section~\ref{sec:selforganizing}. To do so, we run simulations on a platform that integrates real road topologies, realistic vehicle mobility, real maps of obstructions, and empirical signal measurements with 802.11p hardware, all of which are gathered from and applied to the city of Porto, Portugal. The platform, which was custom-developed for this work, is described next.


\subsection{An Urban Simulation Platform Using Real Data} 
\label{sub:simulation_model}
The urban environment introduces a number of challenges to a vehicular network. The movement of cars is dependent on factors such as the time of day, the type of road the car is on, and the activity of intersection traffic lights; the propagation of radio waves is affected by obstructing buildings of varied sizes and composition, signal reflection and diffraction on multiple surfaces, and even by other cars themselves blocking the communication path; moreover, constant node mobility causes these conditions to change at a rapid pace.

To better characterize the complexities of urban areas, we set out to design simulation scenarios that are highly realistic, so that the resulting data is reliable and indicative of what can happen in a real-life scenario. With this goal in mind, we developed a platform with the following features:
\begin{itemize}
	\item Realistic vehicle mobility and traffic light patterns, obtained with the well-known open-source vehicle simulator SUMO~\cite{sumo}.
	\item Real urban street layouts generated from publicly-available city maps that include lane numbers, speed limits and traffic lights~\cite{openstreetmap}.
	\item An accurate, vectorial model of urban obstructions, made available by the Porto City Council. With this obstruction data, in the form of a \emph{shapefile}, we created a Geographic Information System (GIS)~\cite{postgis} to determine Line-Of-Sight status between any two vehicles.
	\item Realistic modeling of core wireless network metrics such as Bandwidth and Packet Loss, as a function of node distance and Line-Of-Sight status, derived from empirical measurements gathered with actual 802.11p-equipped vehicles circulating in Porto as well~\cite{statchannel}.
\end{itemize}

In our simulations, the signal strength between two cars is modeled with the results presented in~\cite{statchannel}. The data from this study are used to assign levels of coverage quality while a parked vehicle is building its coverage map, through the process described earlier in Section~\ref{sub:self_observed_coverage_maps}. This classification criteria is shown in Table~\ref{tab:signalVsDistance}. In a real-life scenario, coverage strength can be determined with received signal strength measurements from the DSRC radios. 

\begin{table}[b]
\centering
\caption{Quality Measure Criteria for Coverage Maps}
\label{tab:signalVsDistance}
\begin{tabular}{@{}clcc@{}}
\textbf{Quality} & & \textbf{Distance (LOS)} & \textbf{Distance (NLOS)} \\ \midrule
5 & & ~70~m & ~58~m \\
4 & & 115~m & ~65~m \\
3 & & 135~m & 105~m \\
2 & & 155~m & 130~m 
\end{tabular}
\end{table}



\subsection{Improving Broadcast Delay in Sparse Networks} 
\label{sub:improving_message_broadcast_delay_in_sparse_networks}
In the initial stage of a vehicular network deployment, insufficient numbers of DSRC-equipped vehicles will cause the network to become sparse. A sparse network is one where some of its nodes are too far apart from their neighbors for communication to occur, leading to network disconnection. In an urban area, this causes virtual clusters of cars to form, where cars in a cluster can talk to one another, but are unable to send messages to other neighboring clusters. This sparse network problem has been well studied on highway vehicular networks~\cite{jsac2007} and, in these environments, connected Roadside Unit deployments bring substantial benefits~\cite{tvt2013,A5}. While this issue is prevalent in scenarios of low market penetration, it can also occur in periods of low traffic, even with full market penetration of DSRC hardware.

Here, we study a scenario where the urban area sees a low density of DSRC-enabled vehicles, which in turn also results in small numbers of parked cars that can join the network as RSUs. Our goal is to determine whether the approaches introduced in this paper bring tangible benefits in these sparse scenarios. For the problem at hand, we simulate an accident that occurs at a random point in the city, automatically triggering the dispatch of an emergency message. In real life, these messages have a dual purpose: to reach any nearby emergency vehicles, hospitals or police stations, and to quickly inform other drivers of the accident, so they can anticipate danger, expect congestion, and take alternative routes. The latter may also aid in the former, by reducing traffic in the vicinity of a crash, which helps emergency services en route.

\begin{figure}[t]
	\centering
	\includegraphics[width=\linewidth]{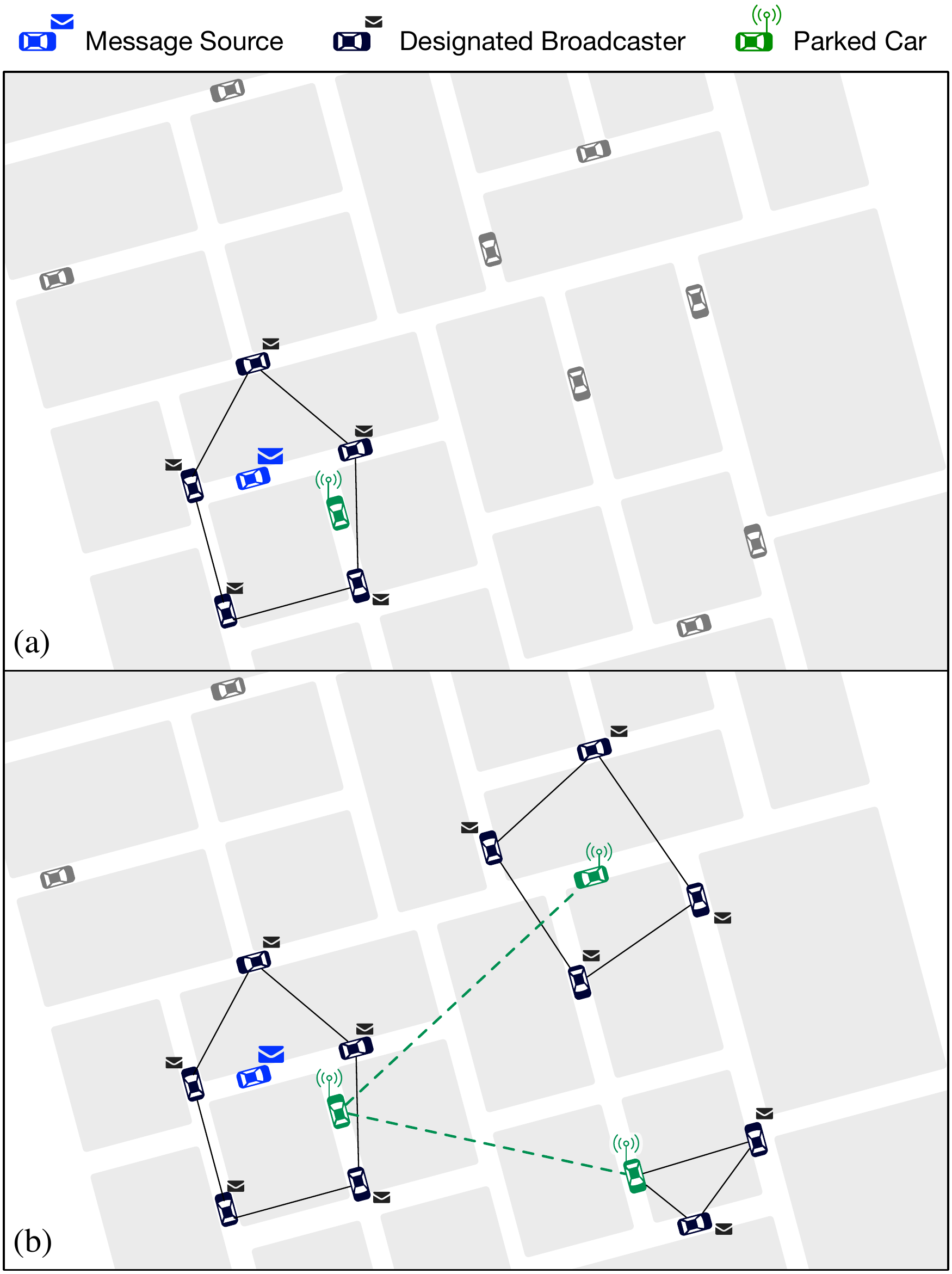}
	\caption{UV-CAST selecting forwarders for a message broadcast with (a)~no RSUs, and (b)~3~parked cars active in standalone mode.}
	\label{fig:uvcast}
\end{figure}

Our metric of interest is then the \emph{message reachability}: the rate at which the message spreads to nearby vehicles, and how quickly everyone in the immediate neighborhood is informed. To broadcast the message, we select UV-CAST, a well-known urban broadcasting protocol designed specifically for vehicular networks~\cite{uvcast}. When the need to disseminate a message occurs, UV-CAST comes into play by locating the edge nodes of the cluster the message is coming from, with a gift-wrapping algorithm. These edge nodes are then designated as the ideal message broadcasters and continue to rebroadcast the message as they meet new vehicles on the road. Figure~\ref{fig:uvcast}a shows an example of the algorithm at work. 

In this analysis, we adapt UV-CAST to support multiple message origins, by making use of the support network of parked cars. An example of this multi-origin system is seen in Figure~\ref{fig:uvcast}b -- here, active parked cars in the area assist the rebroadcasting of messages to other parts of the network. At locations where other active parked cars exist, UV-CAST's algorithms are executed again, and new sets of broadcasters are selected, effectively increasing the numbers of vehicles that redistribute our emergency message, as well as their geographical dispersion.

For simulation purposes, we set up a 1~square kilometer area in the city of Porto, as described earlier in Section~\ref{sub:simulation_model}, featuring a diverse mix of road structures, speed limits, and functional traffic lights. We predefine a ratio of 1 parked car acting as an RSU for every 10 cars on the road (a 1:10 ratio) and analyse three densities of vehicles: 20, 40, and 80 cars per square kilometer. The first two densities match low-density and early-morning scenarios, while the third is a medium-density scenario where network sparsity is expected to be less problematic. Parked cars are placed randomly in the area, and the 1:10 ratio between parked cars and moving cars is chosen to be conservative -- often, close to 80\% of all cars in a given area are parked.

\begin{figure*}[t]
	\centerline{
		\subfloat[]{\includegraphics[width=0.30\linewidth]{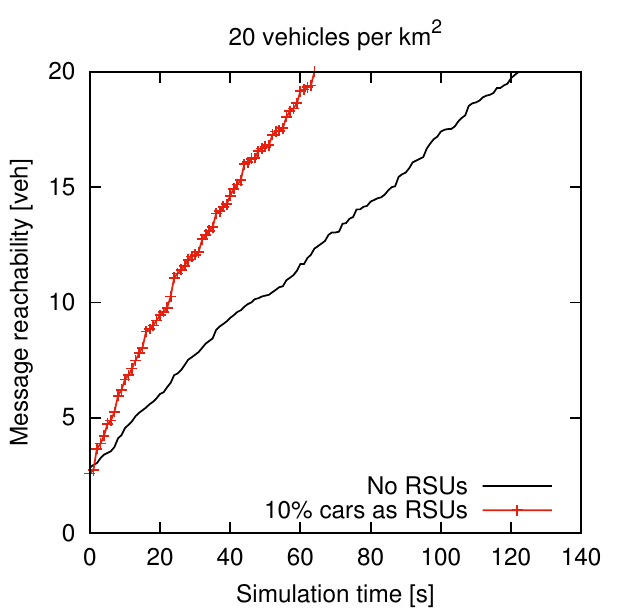}
		\label{fig:rate4cumulative}}
			\hfil
		\subfloat[]{\includegraphics[width=0.30\linewidth]{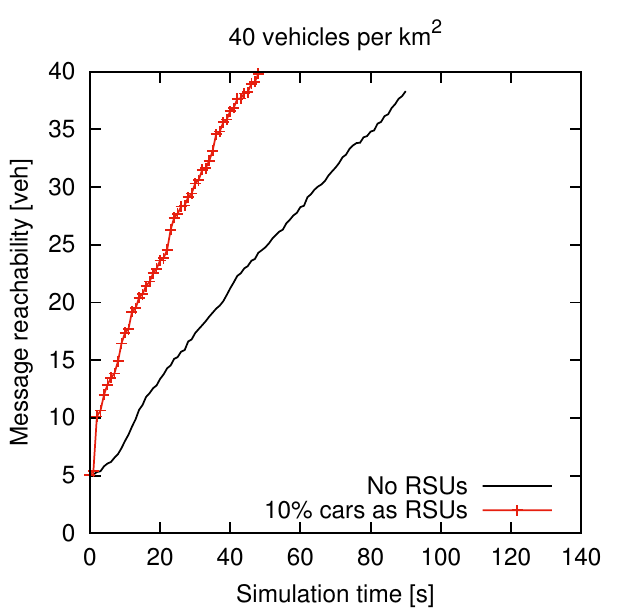}
		\label{fig:rate2cumulative}}
			\hfil
		\subfloat[]{\includegraphics[width=0.30\linewidth]{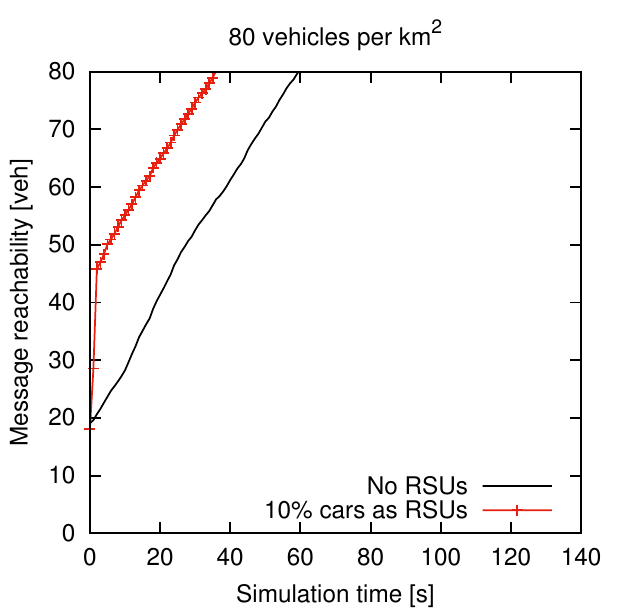}
		\label{fig:rate1cumulative}}
	}
	\caption{Multi-origin UV-CAST message reachability over time, with 10\% of the cars on the road parked and active as standalone RSUs.}
	\label{fig:uvcast-delay}
\end{figure*}

Figure~\ref{fig:uvcast-delay} shows the evolution of an emergency message's \emph{reachability} for the three densities under consideration. Here, reachability denotes the number of nodes that have received the message -- e.g., for the 40~vehicle scenario, the maximum reachability is 40. We compare six scenarios, three with and three without parked car support, and average the results over 50 repetitions each. 
The data show improvements of 47\%, 45\%\, and 41\% in the time required for full reachability, for the 20-, 40- and 80-vehicle densities respectively. These are substantial gains that effectively cut an emergency message’s broadcast delay in half, in these sparse scenarios.

The improvements diminish, proportionately, as the number of cars on the road increases, which is an expected result given that higher vehicle density translates to improved network connectivity (less sparseness). For example, in the medium-density scenario (80~cars), because the network is less sparse, the emergency message immediately reaches $\approx$ 60\% of all cars in under one second. Seeing smaller gains in well-populated networks does not, however, mean that RSUs are not needed when the network is not sparse -- in higher-density scenarios, the benefits of RSUs will come in the form of controlled broadcasting (e.g., preventing \emph{broadcast storms}) and improved network efficiency, among others.

The simulation results shown in Figure~\ref{fig:uvcast-delay} indicate that even a small number of parked cars randomly-distributed throughout an urban area can bring substantial improvements to the broadcast of important emergency messages, which is one of the most important applications of a vehicular network.



\subsection{Performance of Self-Organization with Minimal Data} 
\label{sub:performance_of_self_organization_with_minimal_information}
We now turn our attention to the performance of our self-organizing approach. To do so, we run a series of simulations in a 1~sq.~km. region in the city of Porto using the simulation platform detailed earlier, and analyze the behavior of the algorithms and mechanisms that were shown in Section~\ref{sec:selforganizing}. 

\subsubsection{Time to Build Coverage Maps} 
\label{ssub:time_to_build_coverage_maps}
\begin{figure}[t]
	\centering
	\includegraphics[width=\linewidth]{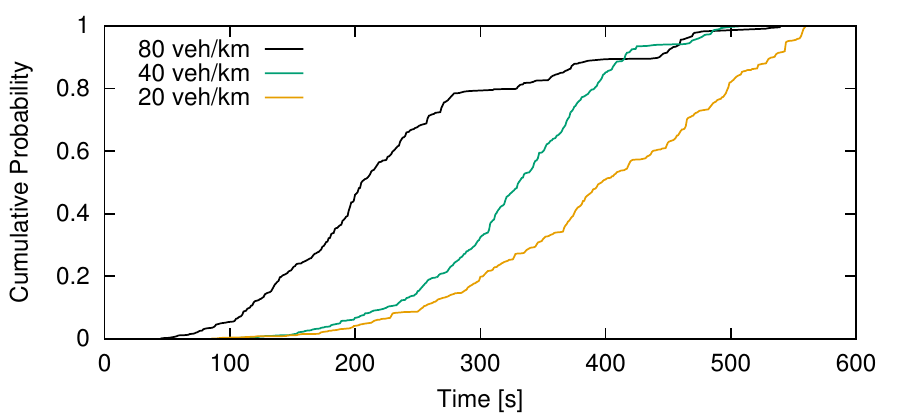}
	\caption{Probability that a car's self coverage map has been completely observed, as a function of elapsed time, in low- and medium-density scenarios.}
	\label{fig:mapTimeCDF}
\end{figure}

The coverage maps that the vehicles build are a fundamental part of the proposed approach, but due to the randomness inherent in a vehicular network, a parked car may not have a reliable way to know if it has overheard enough beacons to form a reliable coverage matrix. A simulator is an ideal tool to analyze this observation step, as it can determine the complete map of a specific car beforehand, and track its evolution. We run a series of simulations where cars are parked at random locations in the city, introduce traffic at three different densities, and track the evolution of each car's coverage map. Figure~\ref{fig:mapTimeCDF} plots the maps' percentage of completeness as a function of time.

The data confirm the intuition that with more moving cars a reliable map will take less time to build. For 80\% of the map to be statistically complete, in these low-density scenarios, the newly parked car must receive beacons for 498, 392, and 330~seconds, for the 20, 40 and 80~vehicle densities, respectively. 
A conservative approach for a real-life scenario is then to instruct cars to build coverage maps for 500 to 600~seconds (i.e., for about 10 minutes) after they park, to ensure a sufficiently complete map. A decision may be executed earlier, particularly if large numbers of beacons are received from multiple directions (suggesting a dense network)~-- in the worst case, the car will temporarily take an unnecessary RSU role, and once more coverage is learned, the decision process is retaken. In higher-density scenarios, our tests show that coverage maps are typically built in 30~seconds to a minute.


\subsubsection{Decision Algorithm vs. an Optimal Solution} 
\label{ssub:decision_algorithm_vs_an_optimal_solution}
Given the cell map division proposed in Section~\ref{sub:self_observed_coverage_maps}, our core metrics for a support network of parked cars are: the mean signal level available to each cell (network coverage); the average number of RSUs covering each cell (network saturation); and a count of how many parked cars take on the RSU role. The optimal solution to a given set of parked cars can be determined by evaluating each possible combination of active and inactive vehicles, with complete knowledge of the candidates and their coverage maps. Here, $2^\text{\#cars}$ scenarios are possible, easily exceeding millions of computations, which makes optimal decisions infeasible in real-life.

We design a simulation scenario where 24 cars are instructed to park randomly in a small 0.18~km\textsuperscript{2} section of the map. This constraint forces the cars to park on nearby streets, so that their coverage overlaps and a decision is made on which cars to keep active. For comparison purposes, we first determine the optimal solution by bruteforcing the $2^{24}$ possible combinations (approx. 16,7 million runs), and we also calculate the effects of simply activating all available parked cars. The results are shown in Table~\ref{tab:algorithmResults}.
\begin{table}[b]
	\renewcommand{\arraystretch}{1.3}
	\centering
	\caption{Reference Optimal Metrics}
	\label{tab:algorithmResults}
\begin{tabular}{@{}lccc@{}}
\textbf{Scenario} & \textbf{Signal Coverage} & \textbf{RSU Saturation} & \textbf{RSU Count} \\ \midrule
All RSUs active & 4.13 & 8.28 & 24 \\
Optimal selection & 4.06 & 1.63 & 5
\end{tabular}
\end{table}
Predictably, the activation of all available vehicles leads to the best coverage, but also causes each cell to see an average of 8 RSUs, which is inefficient and may be problematic. The optimal solution provides roughly the same level of signal coverage, now with only 1.6 RSUs seen at each cell, and just 5 out of the 24 cars left active. These results show that, given perfect decisions, 19 of these 24 parked cars are not needed and can enter an energy-efficient sleep state. 

Next, we analyze our self-organizing approach, with sharing of coverage maps and decisions taken at each node, to determine whether it can approach the outcome of an optimal solution. To do so, we vary the weights of each component in the decision criteria $\mathrm{d_{score}}$, rerun the simulation scenario, and plot the resulting coverage, saturation and RSU count metrics. 

\begin{figure}[t]
	\centering
	\includegraphics[width=\linewidth]{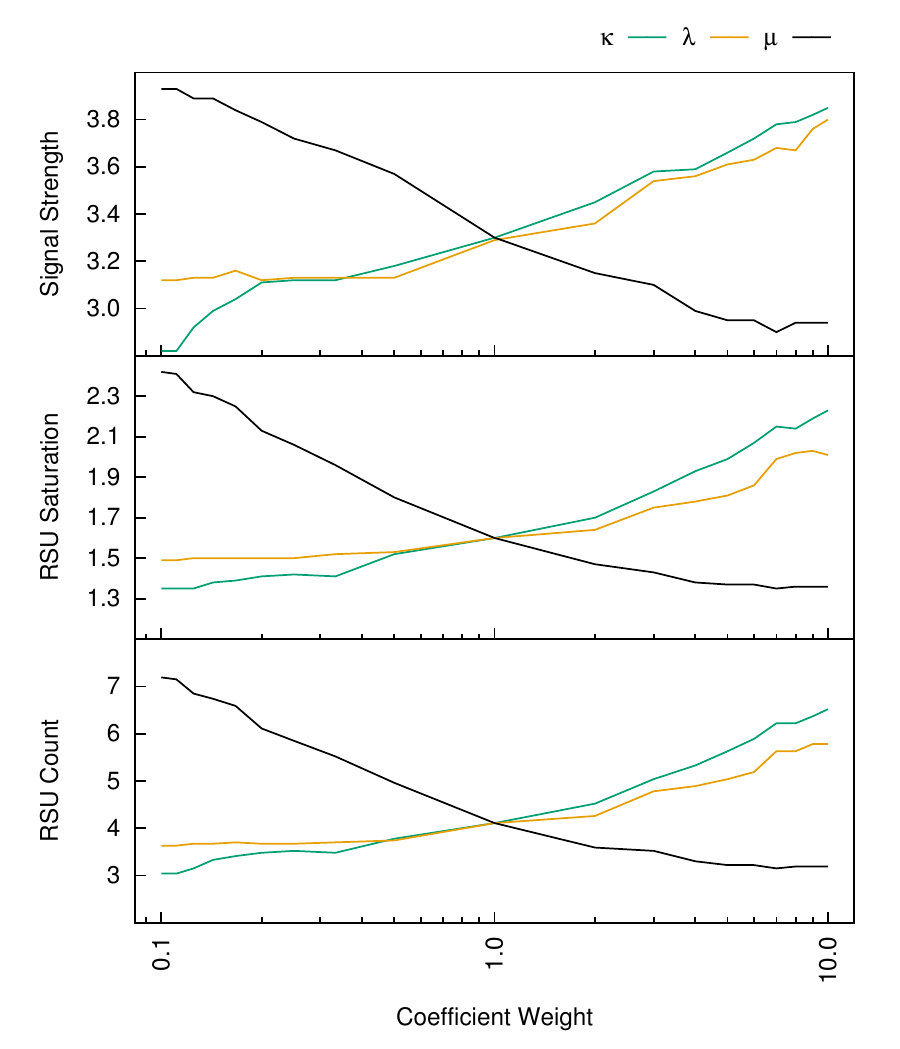}
	\caption{Effect of varying the coefficients $\kappa$, $\lambda$ and $\mu$ on signal strength, RSU saturation, and number of active RSUs.}
	\label{fig:coeff}
\end{figure}

\begin{figure}[t]
	\centering
	\subfloat[]{\includegraphics[width=0.48\linewidth]{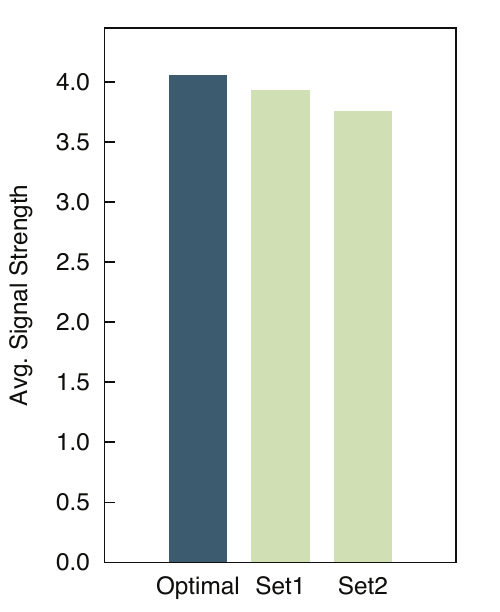}
		\label{fig:boxsignal}}
		\hfil
	\subfloat[]{\includegraphics[width=0.48\linewidth]{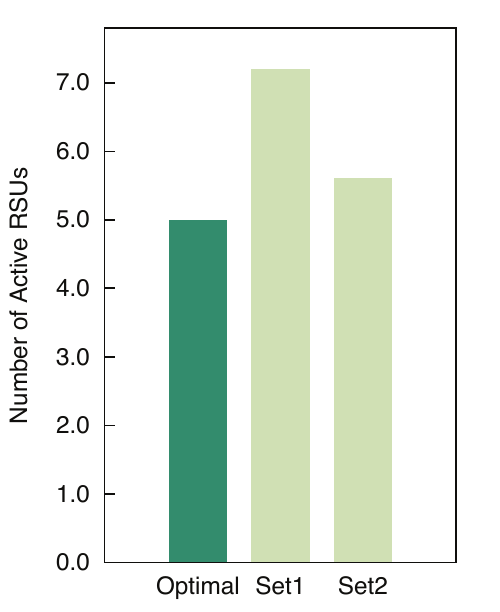}
		\label{fig:boxcount}}
	\caption{(a) Average strength of coverage, and (b) number of parked cars activated, for two coefficient sets with different optimization goals, against an optimal selection. \emph{Set1} attempts to maximize the quality of service coverage, while \emph{Set2} is aimed at reducing the number of active parked cars.}
	\label{fig:boxplots}
\end{figure}

The data in Figures~\ref{fig:coeff} and~\ref{fig:boxplots} show that a decision process optimized towards improved coverage can reach a global signal strength of 3.93, which is only 3\% worse than the Optimal selection. This particular scenario occurs by setting $\mu=0.1$ (the $\mathrm{d_{sat}}$ coefficient). However, the result is also that 7.2 parked cars remain active, which is 2.2 cars more than the optimal solution. A second set of coefficients can be selected to reduce the number of active cars to a minimum: with $\lambda=8$, the resulting signal strength is of 3.76 (7\% worse than Optimal), and only 5.6 parked cars remain active. 

The data presented in this section show that our decision process can obtain near-optimal results without the need for an extensive knowledge of a network's RSUs, and that it can also be directed towards specific goals, such as better signal strength or fewer active parked cars. Extrapolating the results to a 1~sq.~km. area, we observe that about 28-30 parked cars per sq.~km. may be sufficient to form an extensive vehicular support network. This number is an order of magnitude smaller than the typical density of parked cars in a city.



\section{Considerations on Vehicle Battery Life} 
\label{sec:considerations_on_vehicle_battery_life}
The main requirement for a vehicle to work as a Roadside Unit is to be able to maintain power to the DSRC electronics. Ordinary passenger cars have two sources of energy: when the engine is running, an alternator generates electricity, powering the car's electronics and recharging its battery; when the engine is off, power is sourced from that same battery. DSRC electronics, when the car is parked and its engine is off, must be powered from the latter. The amount of energy that is drained should be controlled, as other applications (such as security systems and keyless entry systems) will also be draining the battery at the same time.

Data from our vehicular testbed in the city of Porto, Portugal, using DSRC hardware, show that for the average 6.64~hours that a car spends parked every day~\cite{montreal}, keeping DSRC electronics on for that entire duration will take a \unit[2.8]{\%} toll on a car's battery. Our prototype electronics consume 4.5~watts of power (with radios on a \unit[50]{\%} duty cycle), and production hardware is expected to draw under 3~watts in operation. Using these figures, we reach the \unit[2.8]{\%} figure as follows: vehicular electronics operate at \unit[12]{VDC}, and so the instantaneous current draw of a \unit[3]{W} unit is equal to $I_{radio}  =  P/V = \unit{0.25}{A}$, therefore, the energy consumed over the $\approx$6.64 hours parked is $E  =  I_{radio} \times t_{parked} = \unit[1.66]{Ah}$. On a standard automotive battery, which on passenger vehicles has a capacity of 60~amperes/hour, the amount of energy drained is then ${\%}E_{drain}  =  \unit[1.66]{Ah}/\unit[60]{Ah} = \unit[2.76]{\%}$. Battery capacity degrades with age and usage, and car batteries reach End-Of-Life (EOL) when less than \unit[50]{\%} of the original capacity remains. At this EOL point, the above figure would be doubled, to $\approx$\unit[5.6]{\%}.

Despite these reassuring figures, parked cars acting as RSUs should nevertheless limit their activity, to safeguard excessive drain on the car's battery. With the mechanism shown in Section~\ref{sub:substituting_displaced_cars_in_a_network_friendly_way}, that allows inactive cars to react to displaced RSUs and automatically select a replacement, an active parked car can intentionally disable itself after a prespecified amount of time (e.g., after 1 hour, the impact on the battery will have been of 0.4\%-0.8\%) and be automatically compensated for by neighboring cars. More advanced mechanisms can elect replacements beforehand and execute a soft handoff, however, a detailed study of such mechanisms is beyond the scope of this paper.


\section{Related Work}
\label{sec:relatedwork}
The concept of using cars as Roadside Units started to gain traction as it became apparent that the cost of these units would pose a significant barrier to their deployment. Leveraging moving cars as temporary RSUs in urban areas, by requesting that they make brief stops to aid in emergency message broadcasts, was shown to bring a measurable improvement to such goals~\cite{carsasrsus2013}. Similar concepts, when applied to highway vehicular networks, were able to reduce broadcast delay when opposite-lane traffic was insufficient to relay packets~\cite{icc2014}.
 
The specific idea of \emph{parked cars} as members of the vehicular network is first introduced in~\cite{pvavanet}. This work suggested activating parked cars to increase the number of nodes in a sparse network, improving its connectivity. It reported a 3.3x improvement in node density when 10\% of available parked cars became active, and that 29 parked cars in a 1~km street would be needed to achieve a 100\% connection ratio. These results follow from tenuous assumptions of stable 250~meter radio ranges in the urban area, which our empirical data has shown to be improbable.

The interesting work in~\cite{dressler2014} suggests using parked cars as a means to overcome the signal degradation that occurs when buildings block the line of sight between two vehicles. By activating parked cars at key intersection points, other cars can use the parked car at the intersection as an unobstructed message relay. With this approach, cars were shown to be able to receive nearby emergency messages up to 17~seconds faster. A follow-up to this work~\cite{dressler2012} also demonstrated how parked cars can be used to aid existing RSUs in content downloading. By caching content from fixed RSUs, bandwidth demands on the main RSU for distributing content could be alleviated. These works presented data obtained under the assumption of 200~meter radio ranges, irrespective of obstructions.

The existing body of work on parked cars has revealed the interesting possibilities that can be brought by leveraging parked cars as RSUs, albeit in specific, limited scopes. While these works describe some interesting use cases for parked cars in urban vehicular networks, they do not address the fundamental question of how these entities should be selected, managed, and replaced using a self-organizing network approach, nor whether they can assume more flexible support roles in the network, instead of being relegated to these specific uses. Our work aims to show that using a self-organizing network approach, parked cars can serve as a low-cost and very efficient alternative to deploying RSUs in urban areas.

\section{Conclusion} 
\label{sec:conclusion}
It has been shown that, using a self-organizing network approach, the large numbers of parked vehicles in urban areas can be leveraged to serve as RSUs and provide support to the networks of moving vehicles. These networks of parked cars can serve as low-cost, efficient alternatives to expensive deployments of fixed Roadside Units. To function as RSUs, they require only the ability to keep DSRC electronics powered while vehicles are parked. Our proposed self-organizing network approach introduced novel ways for cars to determine their ability to act as RSUs, and to decide whether to become RSUs from that knowledge. Our analysis showed how a support network that provides excellent coverage is possible using only a small fraction of the cars that are normally parked in a city.




\begin{IEEEbiography}
[{\includegraphics[width=1in,height=1.25in,clip,keepaspectratio]{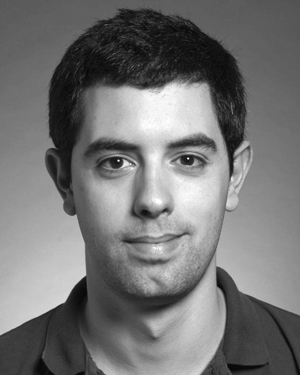}}]{Andre~B.~Reis} is currently working toward the Ph.D. degree in electrical and computer engineering with Carnegie Mellon University (CMU), Pittsburgh, PA, USA, and the University of Aveiro under the CMU-Portugal Program. He received the B.S. and M.Sc. degrees in electronics and telecommunications engineering from the University of Aveiro, Portugal, in 2009, in collaboration with the Eindhoven University of Technology, Netherlands. His current research focuses on infrastructure support systems for vehicular networks in challenging scenarios. He has also published on multimedia Quality of Experience over ad~hoc networks.
\end{IEEEbiography}

\begin{IEEEbiography}
	[{\includegraphics[width=1in,height=1.25in,clip,keepaspectratio]{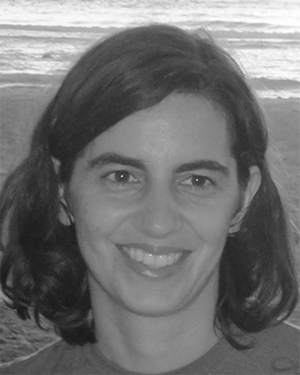}}]{Susana~Sargento} is an Associate Professor with ``Habilitation'' in the University of Aveiro and the Institute of Telecommunications, where she is leading the Network Architectures and Protocols (NAP) group. She has more than 15 years of experience in technical leadership in many national and international projects, and worked closely with telecom operators and OEMs. She has been involved in several FP7 projects, EU Coordinated Support Action 2012-316296 ``FUTURE-CITIES'', national projects, and CMU-Portugal projects. She has been TPC-Chair and organized several international conferences and workshops. She has also been a reviewer of numerous international conferences and journals, such as IEEE Wireless Communications, IEEE Networks, and IEEE Communications. Her main research interests are in the areas of self-organized networks, in ad-hoc and vehicular network mechanisms and protocols, such as routing, mobility, security and delay-tolerant mechanisms, resource management, and content distribution networks. In March 2012, Susana co-founded a vehicular networking company, Veniam, a spin-off of the Universities of Aveiro and Porto, which builds a seamless low-cost vehicle-based internet infrastructure. Susana is the winner of the 2016 EU Prize for Women Innovators.
\end{IEEEbiography}

\vfill\eject

\begin{IEEEbiography}
	[{\includegraphics[width=1in,height=1.25in,clip,keepaspectratio]{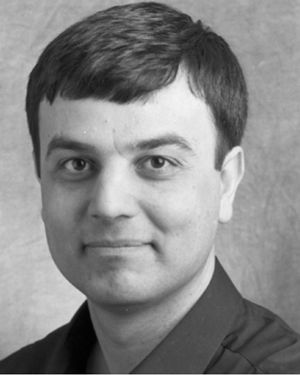}}]{Ozan~K.~Tonguz} is a tenured full professor in the Electrical and Computer Engineering Department of Carnegie Mellon University (CMU). He currently leads substantial research efforts at CMU in the broad areas of telecommunications and networking. He has published about 300 research papers in IEEE journals and conference proceedings in the areas of wireless networking, optical communications, and computer networks. He is the author (with G. Ferrari) of the book \emph{Ad Hoc Wireless Networks: A Communication-Theoretic Perspective (Wiley, 2006)}. He is the inventor of 15 issued or pending patents (12 US patents and 3 international patents). In December 2010, he founded the CMU startup known as Virtual Traffic Lights, LLC, which specializes in providing solutions to acute transportation problems using vehicle-to-vehicle (V2V) and vehicle-to-infrastructure (V2I) communications paradigms. His current research interests include vehicular networks, wireless ad hoc networks, sensor networks, self-organizing networks, artificial intelligence (AI), statistical machine learning, smart grid, bioinformatics, and security. He currently serves or has served as a consultant or expert for several companies, major law firms, and government agencies in the United States, Europe, and Asia.
\end{IEEEbiography}

\vfill

\end{document}